\RequirePackage{fix-cm}
\documentclass[twocolumn,epjc3]{svjour3}  
\smartqed  
\RequirePackage{graphicx}
\RequirePackage[colorlinks,citecolor=blue,urlcolor=blue,linkcolor=blue]{hyperref}
\RequirePackage{amsmath}
\RequirePackage{booktabs}
\RequirePackage{multirow}
\RequirePackage{subcaption}
\RequirePackage{upgreek}
\RequirePackage{xspace}

\newcommand{\decay}[2]{\ensuremath{#1\!\to #2}\xspace} 
\newcommand{\order}[0]{{\ensuremath{\mathcal{O}}}\xspace}

\newcommand{\Pb}[0]{\ensuremath{b}\xspace}
\newcommand{\PB}[0]{\ensuremath{B}\xspace}
\newcommand{\PJ}[0]{\ensuremath{J}\xspace}
\newcommand{\PK}[0]{\ensuremath{K}\xspace}
\newcommand{\Pp}[0]{\ensuremath{\mathrm{p}}\xspace}

\newcommand{\Pmu}[0]{\ensuremath{\mu}\xspace}
\newcommand{\Ppi}[0]{\ensuremath{\uppi}\xspace}
\newcommand{\Ppsi}[0]{\ensuremath{\uppsi}\xspace}   

\newcommand{\pion}[0]{{\ensuremath{\Ppi}}\xspace}
\newcommand{\piz}[0]{{\ensuremath{\pion^0}}\xspace}
\newcommand{\pip}[0]{{\ensuremath{\pion^+}}\xspace}

\newcommand{\kaon}[0]{{\ensuremath{\PK}}\xspace}
\newcommand{\Kp}[0]{{\ensuremath{\kaon^+}}\xspace}

\newcommand{\Kstarz}[0]{{\ensuremath{\kaon^{*0}}}\xspace}

\newcommand{\proton}[0]{{\ensuremath{\Pp}}\xspace}

\newcommand{\bquark}[0]{{\ensuremath{\Pb}}\xspace}

\newcommand{\mup}[0]{{\ensuremath{\Pmu^+}}\xspace}
\newcommand{\mun}[0]{{\ensuremath{\Pmu^-}}\xspace}
\newcommand{\mumu}[0]{{\ensuremath{\Pmu^+\Pmu^-}}\xspace}

\newcommand{\ellell}[0]{{\ensuremath{\ell^+\ell^-}}\xspace}

\newcommand{\jpsi}[0]{{\ensuremath{{\PJ\mskip -3mu/\mskip -2mu\Ppsi}}}\xspace}

\newcommand{\B}[0]{{\ensuremath{\PB}}\xspace}
\newcommand{\Bu}[0]{{\ensuremath{\B^+}}\xspace}

\newcommand{\Bp}[0]{{\ensuremath{\Bu}}\xspace}

\newcommand{\Bz}[0]{{\ensuremath{B^0}}\xspace}

\newcommand{\BToKll}[0]{\decay{\Bp}{\jpsi\left(\ellell\right)\Kp}}
\newcommand{\BToKllrare}[0]{\decay{\Bp}{\Kp\ellell}}

\newcommand{\BToPimumu}[0]{\decay{\Bp}{\jpsi\left(\mumu\right)\pip}}
\newcommand{\BToKmumu}[0]{\decay{\Bp}{\jpsi\left(\mumu\right)\Kp}}

\newcommand{\BToKstllrare}[0]{\decay{\Bz}{\Kstarz\ellell}}

\newcommand{\aunit}[1]{\ensuremath{\text{\,#1}}}

\newcommand{\fb}[0]{\ensuremath{\aunit{fb}}\xspace}
\newcommand{\invfb}[0]{\ensuremath{\fb^{-1}}\xspace}

\newcommand{\cm}[0]{\ensuremath{\aunit{cm}}\xspace}
\newcommand{\hz}[0]{\ensuremath{\aunit{Hz}}\xspace}
\newcommand{\mhz}[0]{\ensuremath{\aunit{MHz}}\xspace}
\newcommand{\second}[0]{\ensuremath{\aunit{s}}\xspace}

\newcommand{\gev}{\aunit{Ge\kern -0.1em V}\xspace}
\newcommand{\gevc}{\ensuremath{\aunit{Ge\kern -0.1em V\!/}c}\xspace}

\newcommand{\mev}{\aunit{Me\kern -0.1em V}\xspace}
\newcommand{\mevc}[0]{\ensuremath{\aunit{Me\kern -0.1em V\!/}c}\xspace}
\newcommand{\mevcc}[0]{\ensuremath{\aunit{Me\kern -0.1em V\!/}c^2}\xspace}

\newcommand{\pt}[0]{\ensuremath{p_{\mathrm{T}}}\xspace}

\newcommand{\cern}[0]{\mbox{CERN}\xspace}
\newcommand{\lhcb}[0]{\mbox{LHCb}\xspace}

\newcommand{\triggercalib}[0]{TriggerCalib\xspace}
\newcommand{\tis}[0] {\ensuremath{\mathrm{TIS}}\xspace}
\newcommand{\tos}[0] {\ensuremath{\mathrm{TOS}}\xspace}
\newcommand{\tistos}[0] {\ensuremath{\mathrm{TISTOS}}\xspace}
\newcommand{\tob}[0] {\ensuremath{\mathrm{TOB}}\xspace}

\newcommand{\hltone}[0]{\texttt{HLT1}\xspace}
\newcommand{\hlttwo}[0]{\texttt{HLT2}\xspace}

\newcommand{\trackmva}[0]{\texttt{HLT1TrackMVA}\xspace}
\newcommand{\twotrackmva}[0]{\texttt{HLT1TwoTrackMVA}\xspace}

\newcommand{\rootpkg}[0]{\texttt{ROOT}\xspace}
\newcommand{\roofit}[0]{\texttt{RooFit}\xspace}
\newcommand{\sweights}[0]{\texttt{sweights}\xspace}
\newcommand{\pythia}[0]{\mbox{\texttt{Pythia}}\xspace}
\newcommand{\evtgen}[0]{\mbox{\texttt{EvtGen}}\xspace}
\newcommand{\photos}[0]{\mbox{\texttt{Photos}}\xspace}
\newcommand{\geant}[0]{\mbox{\texttt{Geant4}}\xspace}

\newcommand{\eg}{\mbox{\itshape e.g.}\xspace}
\newcommand{\ie}{\mbox{\itshape i.e.}\xspace}

\newcommand\numberthis{\addtocounter{equation}{1}\tag{\theequation}}

\journalname{Eur. Phys. J. C}
\begin{document}

\title{A framework and implementation for data-driven trigger efficiency estimation at LHCb 
}


\author{
Johannes Albrecht\thanksref{addr1}
        \and
James Andrew Gooding\thanksref{e1,addr1} \and
Maxim Lysenko\thanksref{addr2} \and
Abhijit Mathad\thanksref{addr3} \and
Alessandro Scarabotto\thanksref{addr1} \and
Tomasz Skwarnicki\thanksref{addr4}
}

\thankstext{e1}{E-mail: jamie.gooding@cern.ch}


\institute{Fakult{\"a}t Physik, Technische Universit{\"a}t Dortmund, Dortmund, Germany \label{addr1}
    \and
Department of Physics, ETH Z{\"u}rich, Z{\"u}rich, Switzerland \label{addr2}
    \and
European Organization for Nuclear Research (CERN), Geneva, Switzerland\label{addr3}
       \and
Syracuse University, Syracuse, NY, United States\label{addr4}
}

\date{Submitted to Eur.~Phys.~J.~C: 8${}^{\rm th}$ April 2026}

\maketitle

\begin{abstract}
Estimations of trigger efficiencies are essential to modern particle physics analyses.
  A data-driven method provides a framework in which to estimate these efficiencies from the properties of reconstructed candidates, described in this paper.
  This paper also presents the design, implementation and performance of a software package, \triggercalib, which provides a first centralised implementation of these calculations and can be seamlessly employed in physics analyses.
  Additionally, the estimation of statistical and systematic uncertainties is discussed.
\end{abstract}

\section{Introduction}
\label{sec:introduction}
Modern particle physics experiments employ multi-stage trigger systems, with the final stage usually performing full event reconstruction~\cite{trigger-review}.
These reconstruct candidates and apply selection criteria to retain only those containing physics of interest to the experiment.
The accurate estimation of the efficiency to select these reconstructed candidates is a crucial part any modern particle physics analysis.
However; these efficiencies cannot simply be evaluated in simulated samples alone, as the complex nature of trigger selections is difficult to model accurately.
To evaluate such efficiencies as the naive fraction of candidates accepted in recorded data would require prohibitively large samples of unfiltered data.
Instead, a data-driven approach, dubbed the \tistos method, is employed to estimate efficiencies of the \lhcb experiment trigger selection using the reconstructed candidates recorded for physics analysis.
This paper presents both the framework of the \tistos method and its first centralised implementation as a Python-based software package.

The \lhcb experiment~\cite{LHCb-DP-2008-001,LHCb-DP-2014-002} is a forward single-arm spectrometer at the LHC, optimised for the study of heavy-flavour hadrons and primarily instrumented in the pseudorapidity range $2 < \eta < 5$.
It collected proton-proton ($\proton\proton$) collision data between 2011 and 2018 during the LHC Run~1-2, reaching an integrated luminosity of 9 \invfb.
Subsequently, it underwent a major upgrade~\cite{LHCb-DP-2022-002} of the detectors and data acquisition system, with the aim of collecting data with a five times greater instantaneous luminosity than the original experiment, reaching ${\mathcal{L} = 2 \times 10^{33} {\cm}^{-2} {\second}^{-1}}$ during the LHC Run~3 from 2022 to 2026.

Prior to the upgrade, \lhcb employed a hardware-based Level 0 (L0) trigger, and a two-stage High Level Trigger (HLT), operated on central processing units (CPU)~\cite{LHCb-TDR-010}.
The hardware trigger was removed in the upgrade, such that a fully software-based trigger is operated in Run 3~\cite{LHCb-TDR-016}.
The first trigger stage (HLT1) operates on graphical processing units, performing a partial event reconstruction at $30\mhz$, whilst the second stage (HLT2) operates on CPUs to perform a full offline-quality event reconstruction at $1\mhz$.
Finally, data are further processed through the offline analysis framework \cite{Mathad:2023zky,Skidmore:2827261}.

The software used in both trigger stages includes a set of algorithms processing information from the \lhcb subdetectors to reconstruct and select processes of interest to the \lhcb physics programme.
The electronic responses of the subdetectors are used for four main reconstruction tasks:
\begin{itemize}
    \item tracking, reconstructing trajectories of particles from hits in the \lhcb subdetectors;
    \item vertexing, searching for the origin locations of $\proton\proton$ collisions or of decaying particles;
    \item particle identification, distinguishing charged final state particles' nature;
    \item neutral particle reconstruction, reconstructing neutral particles, such as photons and \piz mesons, from calorimeter information.
\end{itemize}
The selection algorithms applied to the resulting objects are referred to as trigger ``lines''.

Simulation is required to model the effects of the detector efficiency and the imposed selection requirements.
In the simulation, $\proton\proton$ collisions are generated using  \pythia~\cite{Sjostrand:2007gs,Sjostrand:2006za} with a specific \lhcb configuration~\cite{LHCb-PROC-2010-056}.
Decays of unstable particles are described by \evtgen~\cite{Lange:2001uf}, in which final-state radiation is generated using \photos~\cite{davidson2015photos}.
The interaction of the generated particles with the detector, and its response, are implemented using the \geant toolkit~\cite{Allison:2006ve,Agostinelli:2002hh} as described in \cite{LHCb-PROC-2011-006}. 

This paper provides first an overview of the \tistos method in Sec.~\ref{sec:efficiencies}, which has already been used in many \lhcb physics analyses.
The implementation of this in the \triggercalib software package is then introduced and demonstrated in Sec.~\ref{sec:triggercalib}.
Finally, Sec.~\ref{sec:uncertainties} is dedicated to the estimation of statistical and systematic uncertainties.

\section{The TISTOS method for trigger efficiencies}
\label{sec:efficiencies}

Estimating efficiencies of trigger algorithms in data is non-trivial.
whilst the recorded dataset contains only triggered events, there is sufficient redundancy between different trigger selections and signals to allow the trigger efficiencies to be estimated from these data.\footnote{
A small dataset in which events need not pass a particular trigger is recorded for detector studies, but provides insufficient statistics for the evaluation of trigger efficiencies.
}
Since this redundancy is insufficient to construct an efficiency as the simple fraction of events passing a given trigger selection, the \lhcb collaboration developed a fully data-driven approach to evaluate such efficiencies.
This method, referred to as the \tistos method, was first introduced in \cite{LHCb-DP-2012-004}, is described in ~\cite{LHCb-PUB-2014-039,LHCb-DP-2019-001} and has been used extensively in the \lhcb physics program.
The method is defined in the context of \bquark-decays, though in some cases can be extended to other processes.
Whilst Refs.~\cite{LHCb-DP-2012-004,LHCb-DP-2019-001} discuss the performance of the \lhcb trigger and \cite{LHCb-PUB-2014-039} offers a documentation of the \tistos method, this paper serves as a comprehensive description of the \tistos framework.


\subsection{Trigger categories}
\label{sec:efficiencies/categories}
The \tistos method is built upon a \textit{tag}-and-\textit{probe} approach.
The efficiency of a trigger of interest to select signal candidates is estimated by identifying a subsample of \textit{tag} candidates selected by a certain trigger, assumed to be representative of all candidates in the dataset, and evaluating the fraction of these which are also \textit{probe} candidates, selected by the trigger of interest.
In the \tistos method, \textit{tag} and \textit{probe} categories, named Trigger Independent of Signal (TIS) and Trigger On Signal (TOS), respectively, are defined to construct data-driven trigger efficiency estimators.
These categories characterise individual/aggregate trigger decisions with respect to a user-selected physics-signal candidate, \eg, a reconstructed candidate corresponding to a ${\BToKmumu}$ decay. 
Since trigger decisions are based on finding one or more reconstructed objects satisfying certain trigger criteria, the categorization task reduces to classification of individual trigger-selected objects with respect to each signal candidate selected during the data analysis phase. 
At \hltone level, examples of the selected objects would be those of the \trackmva and \twotrackmva trigger lines.
The former selects significantly displaced high-momentum charged particle tracks; the latter selects displaced high-momentum two-particle vertices.
Similarly, at \hlttwo level an entire ${\BToKmumu}$ candidate may be reconstructed.

Signal candidates and trigger objects both originate from specific electronic signals (``detector hits'') in \lhcb subdetectors, thus the basic \tistos algorithm classifies a set of trigger-related detector hits with respect to a set of detector hits associated with a physics signal candidate.
Composite structure of signal candidates is ignored by collecting together the set of detector hits associated with any constituent part, \eg, only one list of detector hits is created from the three charged tracks constituting a \BToKmumu candidate.
On the other hand, composite structure of trigger objects, if any, is followed.
For example for a dimuon trigger, each muon track is classified independently and the classification must be the same to carry over to the dimuon pair selected by the trigger.
Classification of trigger detector hits with respect to signal detector hits is performed separately for each of the subdetectors playing a key role in reconstruction process, \eg, separately for the Vertex Locator (VELO) \cite{LHCb-TDR-013} and Scintillating Fibre tracker (SciFi) \cite{LHCb-TDR-015} hits.
The set of trigger hits is classified as \tos if sufficiently many of these hits are found among the set of signal hits.
Similarly, the set of trigger hits is \tis if the fraction of trigger hits contained in the set of signal hits is negligible.
If the fraction of trigger hits contained within the set of signal hits is larger than the \tis threshold, but below the \tos threshold, then set of trigger hits does not belong to a useful category for trigger efficiency analysis.
The entire trigger object is \tos (\tis) if all subdetectors used in the classification are \tos (\tis), \eg, for a common background source in which trigger tracks formed of VELO and SciFi hits are \tos for the VELO segment but not for the SciFi segment.

For the most essential tracking subdetectors, VELO and SciFi, the \tis upper bound on the overlapping hit fraction is set to $1\,\%$, while the equivalent lower bound for \tos is set to $70\,\%$, allowing for some imprecision in track reconstruction.
These bounds were tuned on the vertex detector \cite{LHCb-TDR-005} and the outer tracker \cite{LHCb-TDR-006} of the original \lhcb detector, but work well without modification for the equivalent subdetectors of the upgraded \lhcb detector.
Auxiliary tracking subdetectors, the UT \cite{LHCb-TDR-015} and Muon system \cite{LHCb-TDR-004}, are not used for \tos classification since use of the VELO and SciFi is sufficiently robust for the trigger classification scheme.
However, if these subdetector hits were used in constructing the trigger object, they cannot be found among the signal detector hits for the trigger object to be \tis.
ECAL based trigger objects, \eg, photon candidates, are either \tos or \tis depending on whether any ECAL cluster cell is shared with the physics signal side. 
A full trigger line is \tos (\tis) if at least one selected trigger object is \tos (\tis).
Since one trigger line may select more than one trigger object, the same trigger line may be both \tos and \tis at once.
It is also possible to categorise a collection of trigger lines, \eg, the \texttt{Hlt1Global} decision, a logical ``or'' over all \hltone trigger lines, by checking if at least one trigger line is \tos (\tis). 

To facilitate the \tistos method, \hltone and \hlttwo trigger lines persist not only their decision---whether an event passes or fails the trigger criteria---but also detector hits for all selected trigger objects.
The HLT outputs a Decision and Selection Reports covering each trigger lines in a dedicated ``raw bank'' which mimics the data structures output by hardware subdetectors.
\hltone raw bank is added to subdetector raw banks when writing the \hltone output file. 
This allows \hlttwo trigger algorithms to impose \tos (or \tis) criteria on specific \hltone trigger lines.
Particle candidates selected by \hlttwo become the signal with respect to which \hltone trigger objects (lines) are classified.
\hlttwo outputs its own Decision and Selection Reports which can be used for \hlttwo trigger classification with respect to physics signal candidates formed in offline analysis.

Once a specific trigger selection is chosen and classified with respect to each signal candidate, the classification carries over to the candidates, which can be described as:
\begin{itemize}
    \item Triggered on Signal (\tos) type: the signal candidate is sufficient for the given trigger decision, regardless of the rest of the event;
    \item Triggered Independently of Signal (\tis) type: another reconstructed object in the event is sufficient for the given trigger decision, regardless of the signal candidate.
    \item Triggered on Signal and Independently of Signal (\tistos) type: the intersection of the two cases above.
    This is a subset category for both \tos and \tis type candidates. 
\end{itemize}
Additionally, the case where the signal or rest of event alone are insufficient for a trigger decision but their combination is sufficient is dubbed triggered on both (\tob).
A catch-all category for any event which the trigger fires, whether \tis, \tos or \tob,  is also defined, labelled Trig.


Depending on the final state under analysis, if appropriate \hltone and \hlttwo trigger lines target the final state, analysts often impose a \tos requirement on these lines in their offline candidate selection criteria, selecting candidates for which the approach described in Sec.~\ref{sec:efficiencies/tistos} is valid.

If the \tos efficiencies are small for the final state of interest, analysts may include, as a logical ``or'', an explicit \tis requirement among the respective offline selection criteria.
Some analyses may benefit from requiring either \tos or \tis, excluding the \tob category.
Finally, if no explicit trigger criteria are used in the candidate selection, the data-driven approach may still give rather precise results, provided that the \tob category is fractionally small.
This is often the case for studies of final states produced by \bquark-decays: a \tob fraction of $0.5\,\%$ was estimated in \cite{LHCb-PUB-2014-039}.

The assignment of \tis and \tos categories to each candidate in the \lhcb online and offline dataflow is the role of the \texttt{Hlt1TisTosAlg} and \texttt{Hlt2TisTosAlg} algorithms implemented in the DaVinci software application \cite{sprucing-paper}, which classify trigger lines in \hltone and \hlttwo, respectively. 
An interface is introduced through the FunTuple framework \cite{Mathad:2023zky}, in the form of functors \texttt{IS\_TOS} and \texttt{IS\_TIS} which attach the \tis/\tos category to each signal candidate in the output files for a given selection.

\subsection{The TISTOS trigger efficiency method}
\label{sec:efficiencies/tistos}

The trigger efficiency for a given decision can be expressed in terms of the categories defined in the previous section:
\begin{equation}
    \varepsilon_\mathrm{Trig.} = \frac{N_\mathrm{Trig.}}{N_\mathrm{Tot.}} = \frac{N_\mathrm{Trig.}}{N_{\tis}} \times \frac{N_\mathrm{\tis}}{N_\mathrm{Tot.}} = \frac{N_\mathrm{Trig.}}{N_{\tis}} \times \varepsilon_\tis,
    \label{eqn:trig-eff}
\end{equation}
where $N_\mathrm{Trig.}$ is the number of candidates passing the given selection, $N_\mathrm{Tot.}$ the total number of candidates (triggered and not-triggered) and $N_\mathrm{\tis}$ the number of \tis candidates. 
The \tis efficiency, $\varepsilon_\tis$, is not directly measurable in data.
However, $\varepsilon_{\tis}$ can be estimated by the efficiency evaluated within a tag \tos subsample as:
\begin{equation}
    \varepsilon_\tis \approx \varepsilon_{\tis\mid\tos} = \frac{N_\tistos}{N_{\tos}},
    \label{eqn:trig-eff-tis}
\end{equation}
wherein $N_\mathrm{\tistos}$ is the number of candidates which are both \tis and \tos, and $N_\mathrm{\tos}$ is the number of \tos candidates.
This is subject to small correlations which can be easily mitigated by evaluating \tis in a few kinematic bins.
From this approximation, Eq.~(\ref{eqn:trig-eff}) can be written in the form: 
\begin{equation}
    \varepsilon_\mathrm{Trig.} \approx  N_\mathrm{Trig.} \times  \frac{N_\tistos}{N_{\tis}\,N_{\tos}},
    \label{eqn:trig-eff-tistos}
\end{equation}
where all elements can be directly computed in data.
One can arrive at the same formula using the \tos, rather than \tis, category in Eq.~(\ref{eqn:trig-eff}), hence it is symmetrical with respect to these categories.
If the \tis requirement is among the signal selection criteria, then ${N_\mathrm{Trig.}=N_{\tis}}$ and Eq.~(\ref{eqn:trig-eff-tistos}) reduces to ${\varepsilon_\mathrm{Trig.}=\varepsilon_\tis}$ given by Eq.~(\ref{eqn:trig-eff-tis}). Similarly, if \tos is required in signal selection then  Eq.~(\ref{eqn:trig-eff-tistos}) reduces to
$\varepsilon_\mathrm{Trig.}=\varepsilon_\tos$ given by:
\begin{equation}
    \varepsilon_\tos \approx \varepsilon_{\tos\mid\tis} = \frac{N_\tistos}{N_{\tis}}.
    \label{eqn:trig-eff-tos}
\end{equation}
The accuracy to which Eq.~(\ref{eqn:trig-eff-tistos}) reproduces the exact trigger efficiency depends on the quality of the approximations made in Eqs.~(\ref{eqn:trig-eff-tis}) and (\ref{eqn:trig-eff-tos}).
This in turn depends on the signal studied, the selection requirements applied and the choice of trigger lines.  

Correlations between the tag and probe samples in each efficiency may lead to biases in the efficiency evaluation, which often can be studied and corrected for.
In particular, heavy-flavour hadrons are usually produced in pairs in \proton\proton collisions, following the production of heavy-flavour quark and antiquark pairs.
Therefore, the kinematics of the signal candidates may be correlated to other candidates considered independent of signal.\footnote{
    We estimate that candidates for which the \tis arises from another \proton\proton interaction (which are entirely uncorrelated to the \tos signal) account for a fraction of $\order\left(10^{-4}\right)$, and are thus neglected.
}
As trigger selections rely upon requirements on the momentum and impact parameter of the decay products, the assumption of tag and probe independence can induce a small bias in the evaluation of trigger efficiencies when integrating over all phase space.

Additionally, Eq.~(\ref{eqn:trig-eff-tistos}) also applies when a \tos or \tis strategy is chosen, which yields high trigger efficiency in nearly all cases.
The equation can be used without any trigger category selection, \ie, including \tob triggers, and is accurate if, as often is the case, the \tob fraction is small.
This is entirely a mathematical reflection of the fact that any trigger efficiencies studied with tag and probe \tistos method do not probe \tob trigger efficiencies.
As a consequence, each analysis needs to carefully consider trigger selection criteria before employing the method presented here.

\begin{figure}[htb]
    \centering
    \includegraphics[width=\linewidth]{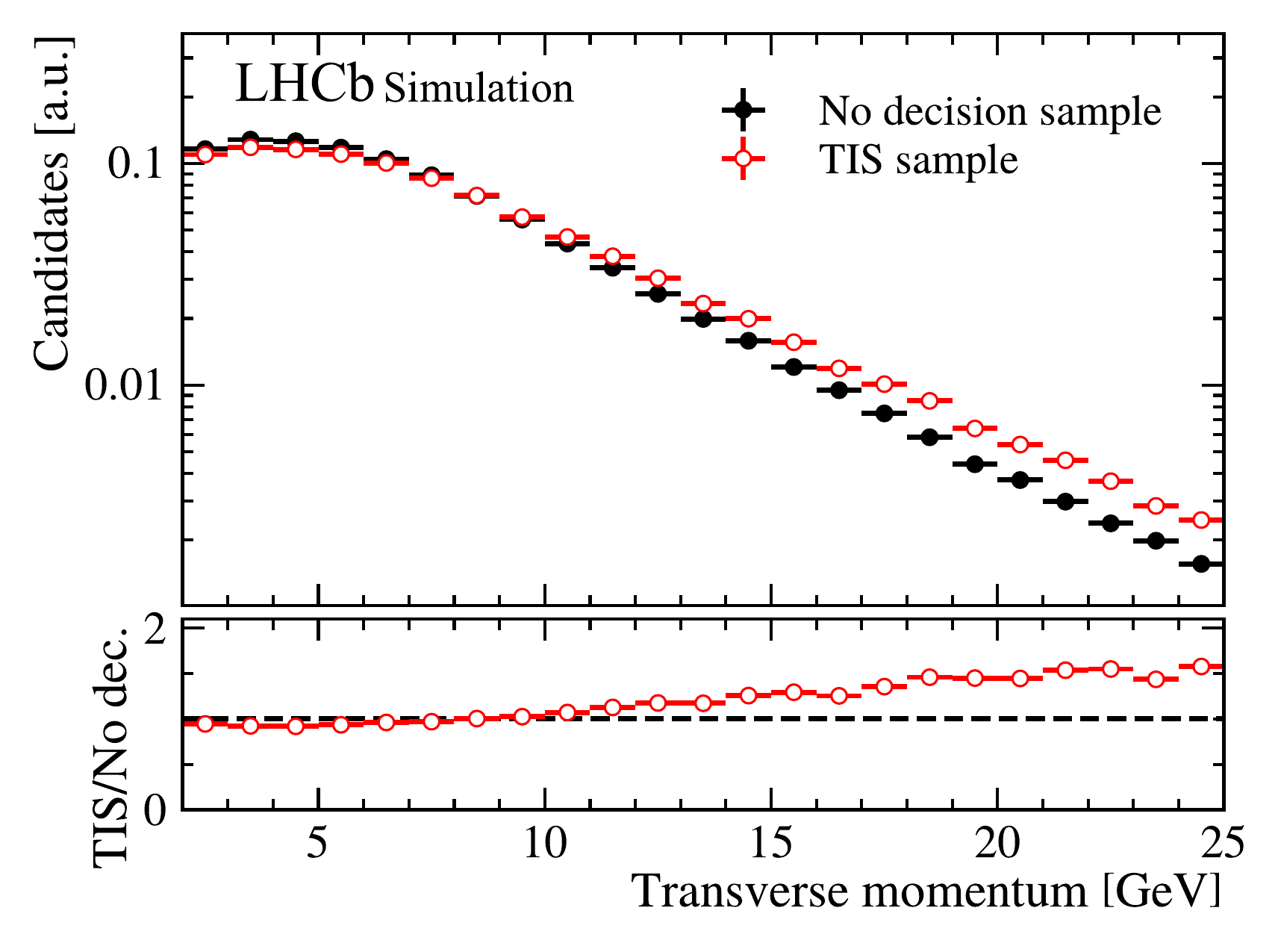}
    \caption{
        Distributions of the \Bp candidate \pt for simulated \BToKmumu decays selected by \tis decisions and without any decisions required.
        It is noted that the \tis sample contains $5.7\times$ fewer events than the sample with no requirement imposed.
    }
    \label{fig:tis_vs_nobias_momentum}
\end{figure}

Kinematic correlations between the \tis and \tos subsamples may bias trigger efficiencies evaluated with the \tistos method.
As shown in Fig.~\ref{fig:tis_vs_nobias_momentum}, the transverse momentum (\pt) of $\Bp$ meson candidates decaying as ${\BToKmumu}$ is harder when selected as \tis candidates than without any specific trigger filtering, overestimating trigger efficiencies.
It is thus necessary to evaluate the efficiency in phase-space intervals, $i$, for which Eq.~\ref{eqn:trig-eff-tistos} can be expressed as
\begin{equation}
    \varepsilon_\mathrm{Trig.} = \frac{N_\mathrm{Trig.}}{\sum\limits_i N_\mathrm{Tot.}^i} = \frac{N_\mathrm{Trig.}}{\sum\limits_i \frac{N_{\tis}^i}{\varepsilon_\tis^i}} =  \frac{N_\mathrm{Trig.}}{\sum\limits_i \frac{N_{\tis}^i N_{\tos}^i}{N_{\tistos}^i}}.
    \label{eqn:trig-eff-binned}
\end{equation}
As the intervals are made more narrow, \ie, the number intervals increases, in each dimension, the trigger efficiencies estimated with the \tistos method become largely unbiased to the correlations between the \tis and \tos candidates, with the estimated efficiencies approaching the fraction of events passing the selection evaluated in a truth-level simulated sample~\cite{LHCb-PUB-2014-039}.
A variety of kinematic variables should be considered when evaluating trigger efficiencies, such as the momentum, transverse momentum, pseudorapidity, lifetime or event occupancy of the candidates studied.
The \triggercalib package provides analysts the flexibility of choosing any variable and phase-space division scheme.

In \lhcb analyses, the \tistos method is applied in two different approaches, reflecting the relevant physics case.
The first approach evaluates trigger efficiencies directly in the data samples studied in the analysis and is only possible with sufficiently large samples.
For processes in which insufficient statistics are available, \eg, rare decay processes such as \BToKllrare, this approach is not possible.
Instead, the efficiencies must be calibrated using a well-understood high-statistic control channel which is kinematically and topologically similar.
To calibrate the trigger efficiencies in the sample of interest, the efficiencies of the control mode must be evaluated in data and in simulated samples, taking the ratios of these as per-event weights to apply to simulated samples of the channel of interest.
For example, the latter approach was used in the \lhcb measurement of the lepton flavour universality ratios $R_K$ and $R_{K^\ast}$ in \BToKllrare and \BToKstllrare decays \cite{LHCb-PAPER-2022-045,LHCb-PAPER-2022-046} 
where the \BToKll decay was used as control channel.

\section{The TriggerCalib software package}
\label{sec:triggercalib}

The \triggercalib software package implements the calculations of the \tistos method as a set of Python-based tools for use in analyses of \lhcb data.
Previously, these calculations were implemented anew in most analyses, requiring analysts to dedicate time and effort to carefully incorporating the \tistos calculations, often needing trigger-specific expertise to do so.
In the first analyses using \triggercalib, analysts have configured the tool in $\order({\rm minutes})$, whereas reimplementing the calculations typically takes $\order({\rm days})$.
This first centralised implementation of the calculations circumvents this, streamlining both the development and validation of physics analyses.

The package is made available through the \cern GitLab at \cite{triggercalib-gitlab} and is deployed to PyPI \cite{triggercalib-pypi}.
A comprehensive documentation of the package is provided at \cite{triggercalib-docs}, containing a user guide, tutorials and a full reference of the \triggercalib code.
The package is built upon functionality of the \rootpkg data analysis framework \cite{Brun:1997pa}, producing \rootpkg histogram and graph objects which analysts can manipulate directly or save for later use, \eg, to take the ratio of efficiencies to produce corrections to simulation.
Analysts must therefore only familiarise themselves with the interface of the tool, since it accepts inputs and provides outputs which are familiar within the \lhcb offline analysis ecosystem.

Two aspects are proposed and implemented in the package, which reflect the different approaches required by different analyses:
\begin{itemize}
    \item data-driven efficiency estimation: evaluation of trigger efficiencies directly on data with the \tistos method, namely on channels with high statistics and well-understood backgrounds;
    \item corrections to simulation: evaluation of per-candidate corrections which can be used to correct the trigger response in simulated samples. For example, in decay modes where the statistics are insufficient for a direct evaluation of the trigger efficiencies, corrections can be computed in a higher statistics reference channel and applied to simulation representative of the target channel.
\end{itemize}
The choice of approach depends on the type of analysis performed, the signal channel studied and the trigger selection(s) of interest.
For sufficiently high statistics channels, the direct evaluation of efficiencies on data should be chosen, whilst the data-simulation correction approach should be used when studying rare processes with limited signal yields.
Correction weights can be evaluated on an abundant control sample similar in kinematics and topology to the signal channel and then applied to simulated samples of the signal decay.
With this approach, the trigger efficiencies can then be computed directly on the weighted simulated sample.
In data, background contributions must be accounted for, applying background mitigation methods to retrieve the necessary \tistos yields for signal alone.


To obtain reliable trigger efficiencies for a given channel, the yields of the \tistos method must contain a negligible contribution from background(s).
Background contributions are mitigated by either subtracting the corresponding amount of background or direct modelling the background component(s), distinguishing signal from background according to a chosen discriminating variable.
Three methods of background mitigation are implemented in \triggercalib:
\begin{itemize}
    \item Sideband subtraction: signal and sideband windows are defined in the discriminating variable.
    In each phase-space region, the density of candidates in the sideband window is taken as an estimate of the density of background candidates within the signal window and subtracted accordingly;
    \item Fit-and-count: a statistical model with components describing signal and background contributions is fit to the discriminating variable for each phase-space region.
    The yields of the signal component in the model are then used to evaluate the efficiency;
    \item \textit{sPlot}: a statistical model is fit globally, \ie, integrated over the other phase-space dimensions, to the discriminating variable.
    Per-event \textit{sWeights} are calculated for each component of the fit according to the \textit{sPlot} method~\cite{Pivk:2004ty}.
    Signal \textit{sWeights} are applied to the distribution(s) of interest, which are then summed in each phase-space region to evaluate the \tistos yields.
\end{itemize}

The sideband subtraction method provides a simple and robust approach across the phase space, even for regions containing relatively few candidates.
However, it is valid only when a linear description of the background component(s) is possible.
When studying candidates in datasets with more complex backgrounds, the fit-and-count and \textit{sPlot} methods must be used to reliably estimate trigger efficiencies.

The three methods are demonstrated in a simulated sample of \BToKmumu decays, reflecting the data-taking conditions of 2024.
This decay channel is widely used as a control channel in \lhcb analyses and is characterised by high statistics and purity: in Run 3, these decays are selected at a rate of $\order\left(10\hz\right)$.
Candidates were selected according to the nominal \hltone configuration in place in 2024, a dedicated \hlttwo line for \BToKmumu decays, and further selection cuts on kinematic, topological and particle-identification properties applied offline.
This sample is supplemented by an artificial combinatorial background component, described in App.~\ref{sec:combinatorial}.
The discriminating variable used corresponds to the invariant mass of \Bp candidates, evaluated as the combination of the \jpsi and \Kp.

These demonstrations estimate the combined efficiency of \trackmva and \twotrackmva, as introduced in Sec.~\ref{sec:efficiencies/categories}.
These two trigger decisions are used by many LHCb analyses and therefore constitute a suitable benchmark.
The combination of these is used to define both the \tos and \tis categories.




A signal window, $\Delta_\mathrm{signal}$, must be defined in the discriminating variable, containing $N_{\rm signal}$ candidates around the signal distribution. 
Outside of $\Delta_{\rm signal}$, sideband windows ($\Delta_\mathrm{sideband}^i$) containing $N_{\rm sideband}^i$ candidates can be defined which contain only the underlying background, from which a background density ($\rho$) is defined.
The background-subtracted signal yield is then obtained as
\begin{equation}
    N_\mathrm{signal}^\prime = N_\mathrm{total} - \Delta_\mathrm{signal} \cdot\rho ~~\mathrm{where}~~\rho = \frac{\sum\limits_{i} N^{i}_\mathrm{sideband}}{\sum\limits_{i} \Delta^{i}_\mathrm{sideband}}.
    \label{eq:sideband_sub}
\end{equation}
This holds for all approximately constant backgrounds, and more generally holds so long as the sideband windows are chosen carefully to reflect the shape of the background.
The method can then be applied per phase-space region to evaluate the \tis-, \tos- and \tistos-filtered yields and estimate the trigger efficiency per Eq.~\ref{eqn:trig-eff-binned}.

\begin{figure}[htb]
    \centering
    \begin{subfigure}[t]{\linewidth}
        \centering
        \includegraphics[width=\linewidth]{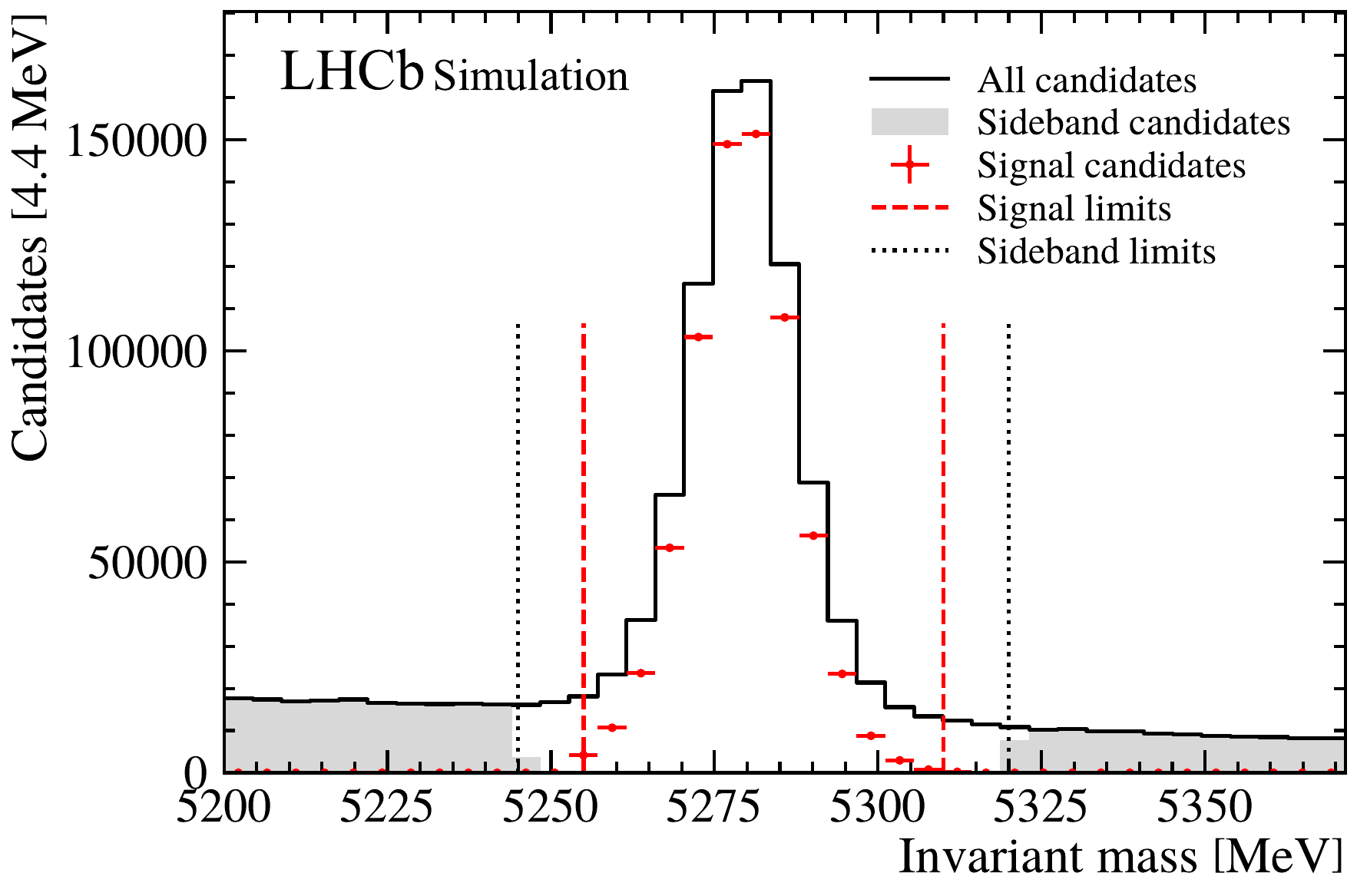}
    \end{subfigure}
    \begin{subfigure}[t]{\linewidth}
        \centering
        \includegraphics[width=\linewidth]{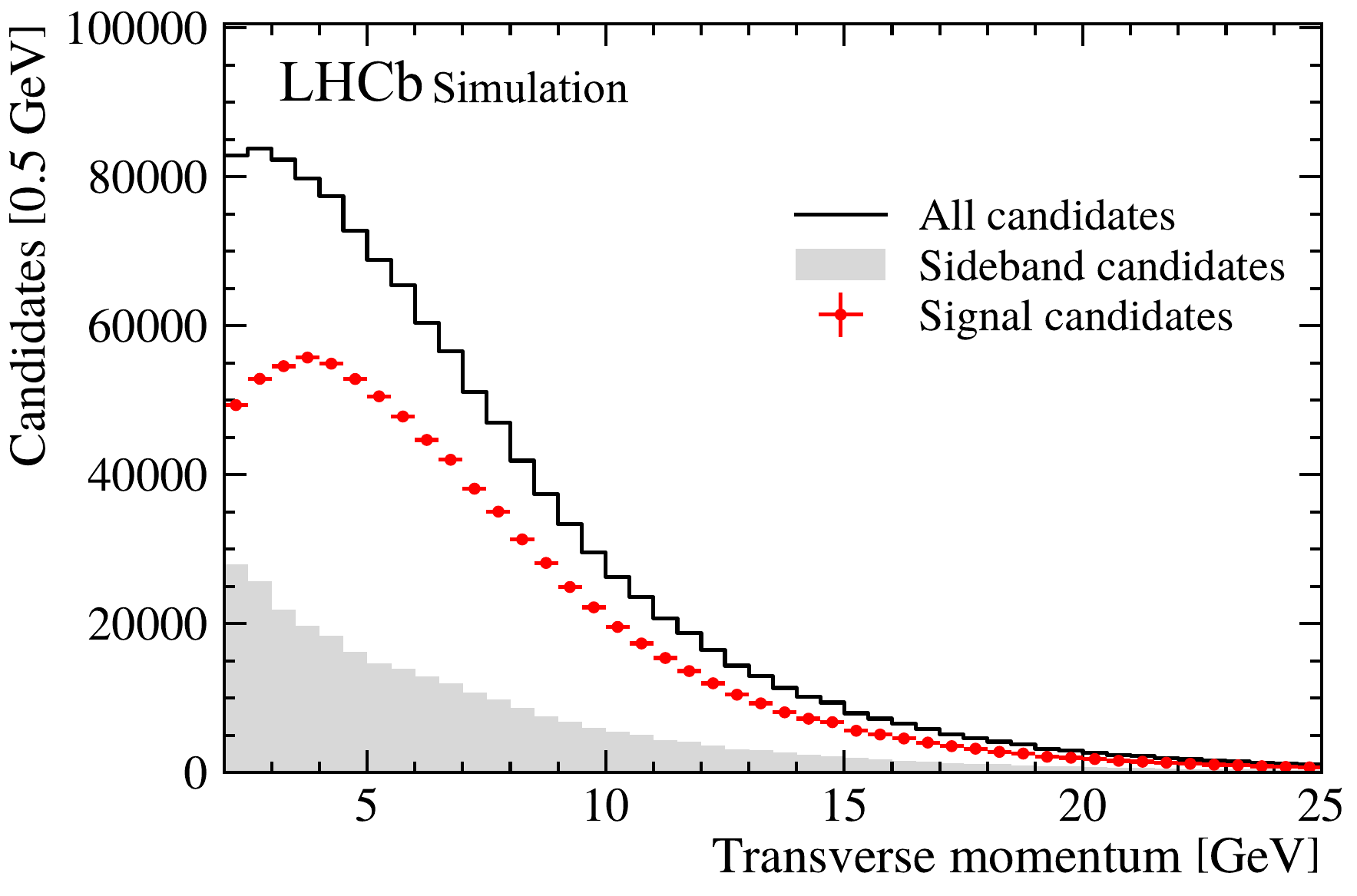}
    \end{subfigure}
    \caption{
        Distributions of candidate (above) invariant mass and (below) transverse momentum for the ${\BToKmumu}$ decay sample.
        Signal and sideband regions used for sideband-subtraction are shown as red and black lines, respectively.
        Candidates falling within the sidebands are shown in grey, whilst the sideband-subtracted candidates are shown in red.}
    \label{fig:sideband-sub-effect}
\end{figure}

In the demonstration, a signal window is defined as ${m\left(\jpsi\Kp\right) \in \left[5255, 5310\right]\mevcc}$, \ie, centred on the known \Bp mass~\cite{PDG2022}.
Two sideband windows are considered: the first for ${m\left(\jpsi\Kp\right) \in \left[5200, 5245\right]\mevcc}$; the second for ${m\left(\jpsi\Kp\right) \in \left[5320, 5375\right]\mevcc}$.
The background is mainly composed of random combinations of tracks which are reconstructed and selected as candidates, so-called combinatorial background.
For simplicity, misidentified background contributions from ${\BToPimumu}$ are not considered.
The signal and sideband windows are shown in Fig.~\ref{fig:sideband-sub-effect}, along with the sideband-subtracted distribution of candidates.


The fit-and-count method exploits extended probability density functions (PDFs) to describe the discriminating variable distribution, to which a negative log-likelihood (NLL) fit is performed in each considered region.
This approach allows analysts to define separate PDFs describing the signal and background components, forming an extended sum of these.
As such, the fit-and-count method appropriately handles more complex background contributions.
However, this approach relies on the likelihood fit converging stably in all regions, which may not be true in phase-space regions with few candidates.
In \triggercalib, likelihood fits using the \roofit~\cite{Verkerke:2003ir} fitting library commonly used in \lhcb are supported.

\begin{figure}[htb]
    \centering
    \includegraphics[width=\linewidth]{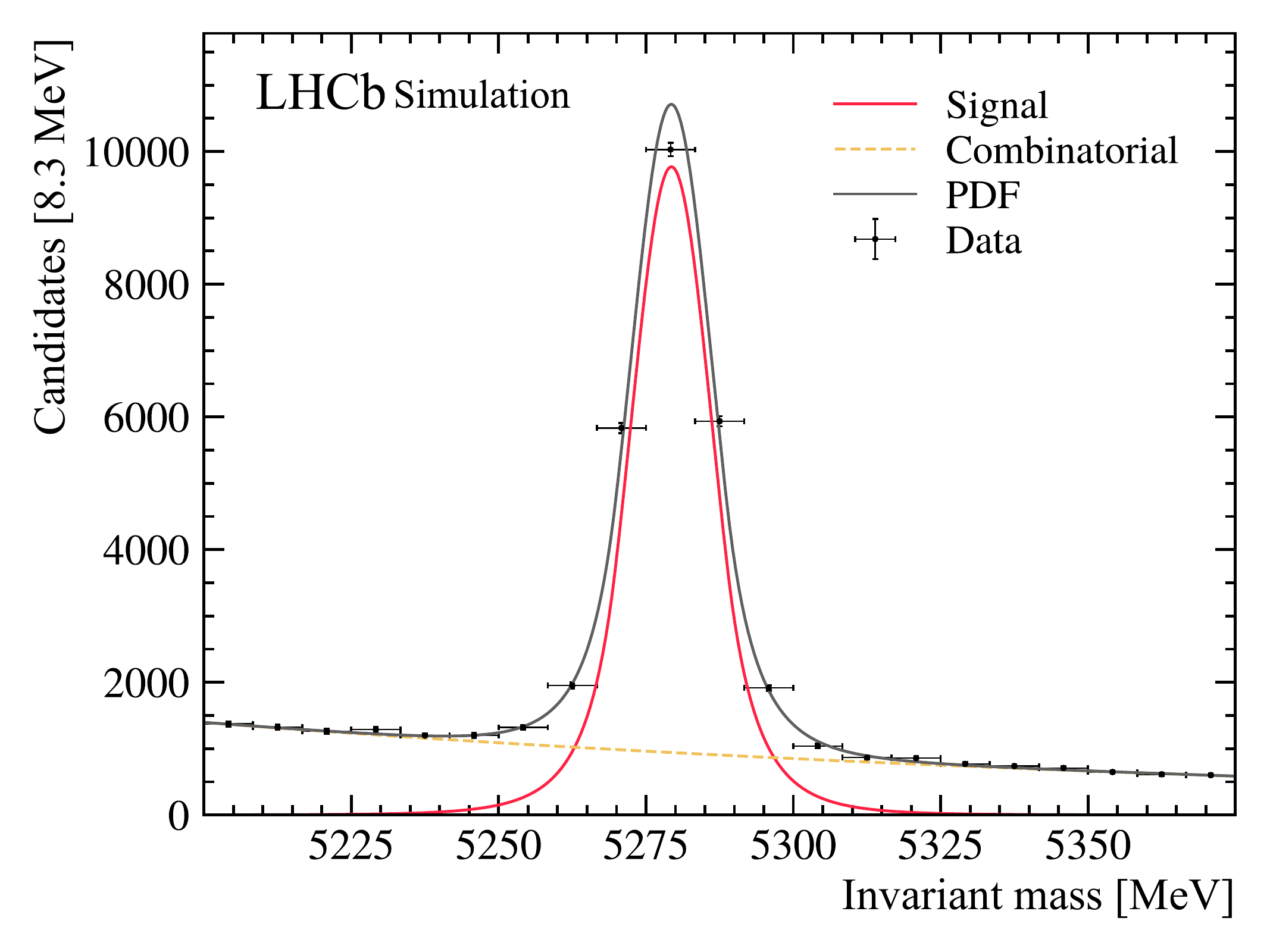}
    \caption{
    Distribution of the \jpsi\Kp invariant mass distribution for a single \pt bin, overlaid with the results of the likelihood fit.
    }
    \label{fig:fit-and-count}
\end{figure}

For demonstration, signal and combinatorial background contributions are described by a double-sided Crystal Ball function~\cite{Skwarnicki:1986xj} and an exponential distribution, respectively.
Parameters describing the signal are obtained from likelihood fits to truth-matched simulated samples, with the mean and width of the distribution varying in fits to the full simulated sample with added background.
The mass distribution for one example \pt bin is shown in Fig.~\ref{fig:fit-and-count}, overlaid with the likelihood fit result.

\begin{figure}[htb]
    \centering
    \includegraphics[width=\linewidth]{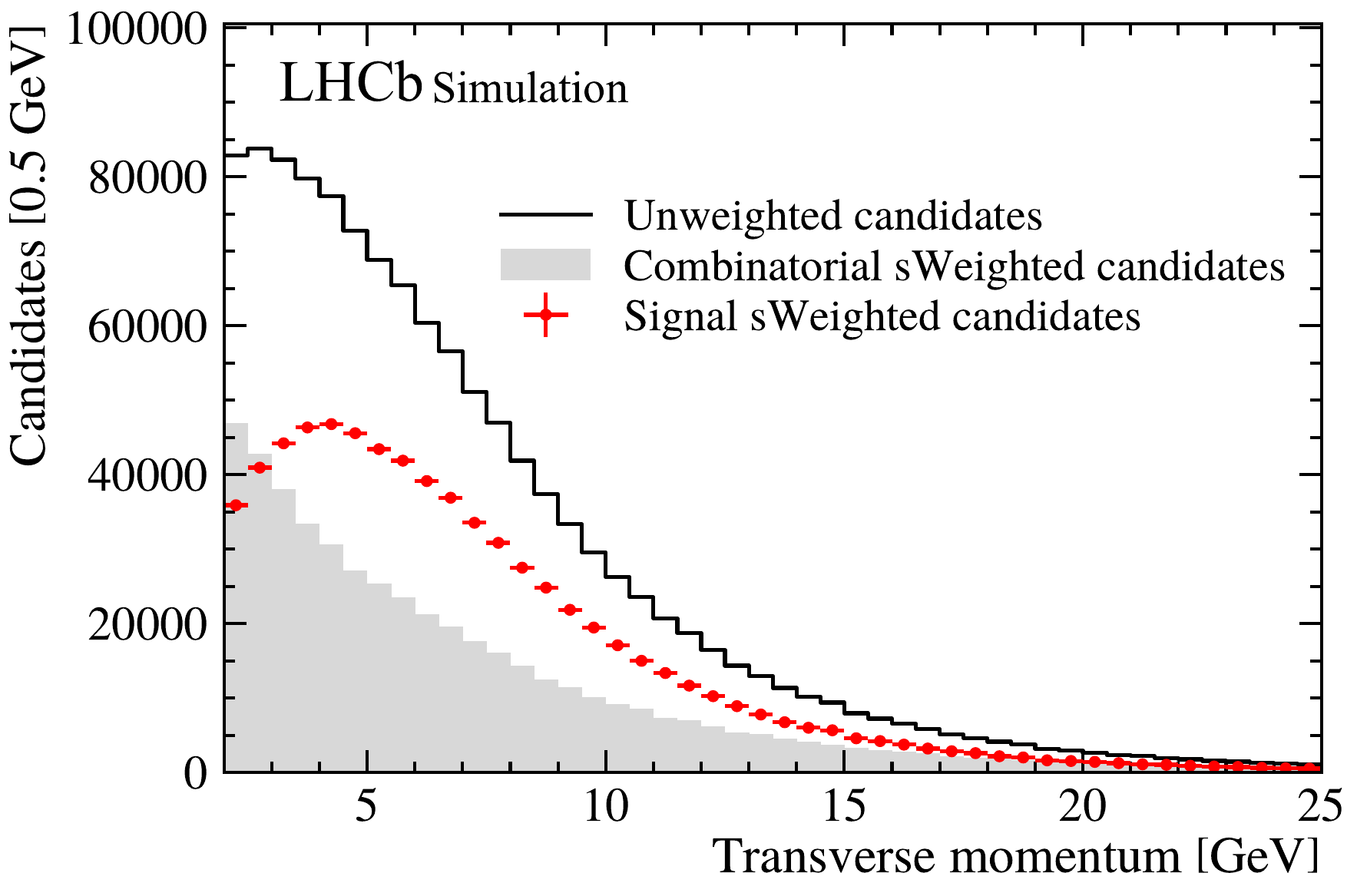}
    \caption{
        Distributions of the \Bp transverse momentum with \textit{sWeights} of the signal and combinatorial background components applied.
        These \textit{sWeights} are calculated from a single likelihood fit to the entire ${\BToKmumu}$ sample. 
    }
    \label{fig:sweighted-distributions}
\end{figure}

As in the fit-and-count method, the \textit{sPlot} method uses an extended sum of PDFs describing signal and background components.
However, rather than fitting in every phase-space region, a global likelihood fit is performed to compute signal \textit{sWeights} in the \textit{sPlot} formalism~\cite{Pivk:2004ty}.
The number of signal candidates in each $i$-th region of the phase space can then be evaluated as the per-region sum of signal \textit{sWeights}, $w_j^i$ for each candidate $j$ in region $i$.
The distributions of $\pt\left(\Kp\mup\mun\right)$ weighted according to signal and background \textit{sWeights} are shown in Fig.~\ref{fig:sweighted-distributions}.

The advantage of this approach is that only a single likelihood fit per category is sufficient to obtain all of the information required to evaluate the trigger efficiencies with the \tistos method in regions of the phase space.
This is in contrast to the fit-and-count method, in which many fits must be performed per category.
However, the \textit{sPlot} formalism is only valid if the variables of interest are uncorrelated to the discriminating variable.


Correlations between control and discriminating variables could lead to a biased evaluation of the trigger efficiency.
Two tests of this, proposed by the \sweights package~\cite{Dembinski:2021kim}, are implemented in \triggercalib: the likelihood ratio test and Kendall's $\tau$ test.
Both tests are discussed in detail and performed in App.~\ref{sec:factorisation}.
Whilst the conclusion of these tests in the sample studied is that the two variables are not independent, if the bias in the \textit{sWeights} affects each category (\tis, \tos, etc.) equally, then these effects may cancel in the estimated efficiencies.



The sideband subtraction, fit-and-count and \textit{sPlot} methods are compared: integrated trigger efficiencies are evaluated according to Eq.~\ref{eqn:trig-eff-binned}, dividing the phase space into regions of the \Bp transverse and longitudinal momenta, and listed in Table~\ref{tab:effs_validity}.
The phase space is divided into 5 bins of $\pt\left(\Kp\mup\mun\right) \in \left[2, 25\right]\gevc$. 
Bins are chosen to contain an approximately equal number of candidates in the \tistos category, as this is the smallest component of Eq.~\ref{eqn:trig-eff-binned}.
All three efficiencies are consistent with one another, as expected for a well-understood control channel with well-behaved backgrounds; however, this is not necessarily generally true.
The difference in sensitivity between the sideband subtraction and fit-and-count/\textit{sPlot} methods arises entirely from uncertainties in fitting, and hence is dependent on the construction and quality of the fits to the discriminating variable.
\begin{table}[htb]
    \centering
    \caption{
        Trigger efficiency evaluated  with the \tistos method for the ${\BToKmumu}$ simulated sample.
        Different background mitigation approaches are compared.
        The \tistos method is applied splitting the sample regions of \pt  and $p_z$ of the $B$ mesons.
        Accompanying uncertainties are purely statistical.
    }
    \begin{tabular}{l c }
        \toprule
        Background mitigation   & Trigger efficiency
        / $\%$   \\
\midrule
 Sideband subtraction     & $97.328 \pm 0.054$                                                                \\
Fit-and-count  & $97.31 \pm 0.11$  \\
\textit{sPlot} & $97.314 \pm 0.094$  \\
    \bottomrule
    \end{tabular}
    \label{tab:effs_validity}
\end{table}

\begin{figure}[htb]
    \centering
    \includegraphics[width=\linewidth]{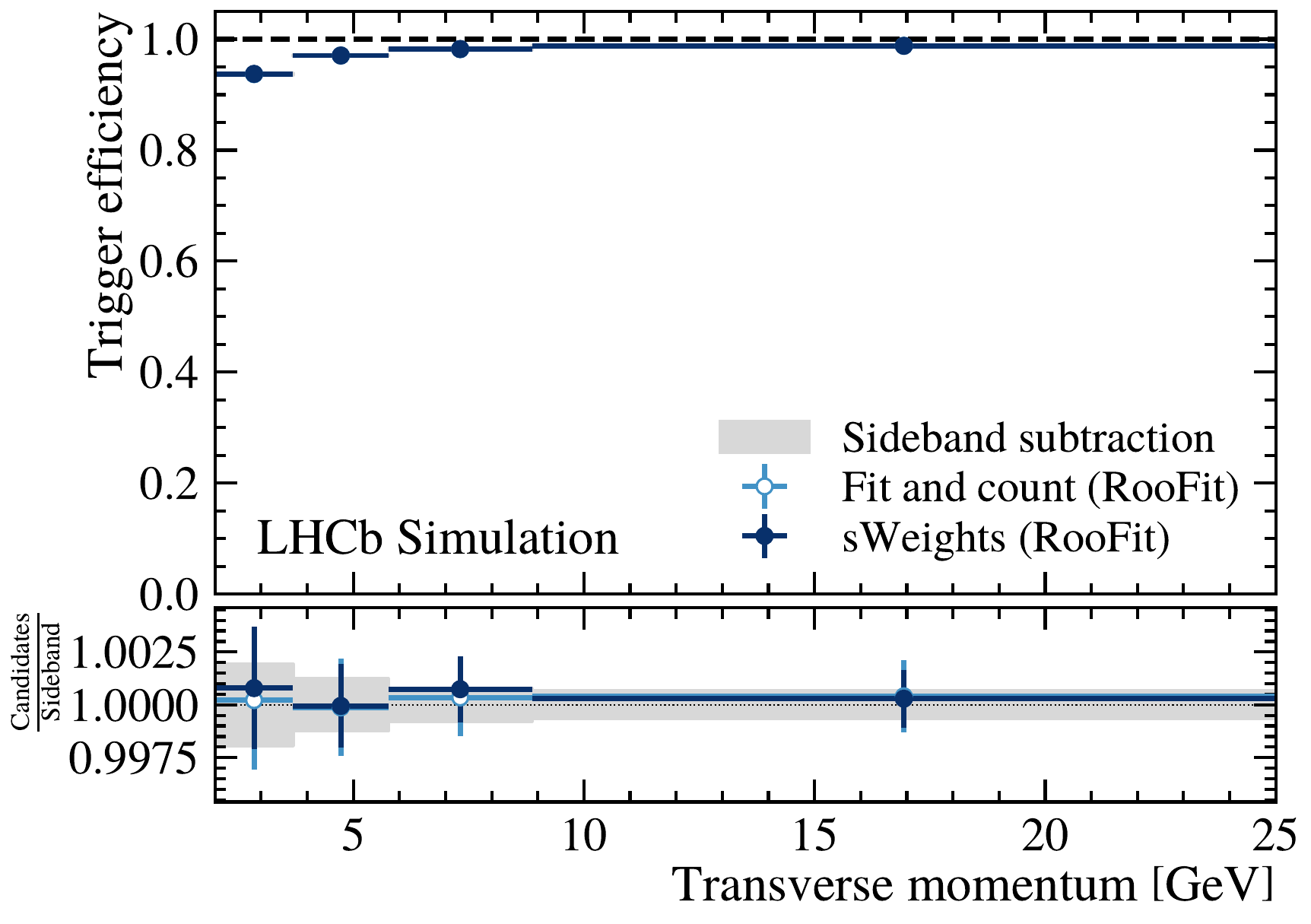}
    \caption{
       Trigger efficiencies evaluated on the ${\BToKmumu}$ simulated sample as a function of the \pt of the $B^+$ meson candidate.
       The different background mitigation methods implemented in the \triggercalib package are compared and return consistent results.
    }
    \label{fig:methods-vs-pt}
\end{figure}

An additional comparison is made for the trigger efficiencies as a function of \pt, shown for a 1-dimensional binning of the sample in four $\pt$ bins in Fig.~\ref{fig:methods-vs-pt}.
As for the integrated trigger efficiencies, the trigger efficiencies in each bin are consistent for each of the three background mitigation approaches.


The \triggercalib software package allows the calculation of correction weights to the trigger response of simulated samples.
The correction weights for efficiencies computed in phase-space bins $i$,\footnote{
Such a binning scheme can be of an arbitrary number of dimensions, though 1- and 2-dimensional phase-space binning schemes are supported by \triggercalib.
} are given by 
\begin{equation}
    w_i = \frac{\varepsilon^\mathrm{data}_i}{\varepsilon^\mathrm{simulation}_i},
    \label{eq:data-mc-weights}
\end{equation}
with $\varepsilon^\mathrm{data/simulation}_i$ as efficiencies computed with the \tistos method on a data/simulated sample in a phase-space region $i$.

\section{Uncertainties}
\label{sec:uncertainties}

The accurate estimation of statistical and systematic uncertainties on trigger efficiencies requires careful consideration of the \tistos method, its categories and its underlying assumptions.
Statistical uncertainties must account for the overlap which appears, by construction, between the different categories. 
As this is non-trivial to derive, per Sec.~\ref{sec:uncertainties/statistical}, and to implement for each use of the \tistos method, an implementation of this is included in \triggercalib.
Systematic uncertainties must account for sources of bias in the estimation of efficiencies, which are often analysis- and implementation-dependent.
These are largely trivial to evaluate as they involve performing the calculations with a modification to the configuration, \eg, to the choice of kinematic variables, thus no functionality is provided for this in \triggercalib and a guide to common sources of systematic uncertainty and their estimation is given in Sec.~\ref{sec:uncertainties/systematic}.

\subsection{Statistical uncertainties}
\label{sec:uncertainties/statistical}

Statistical uncertainties on trigger efficiencies estimated with the \tistos method are typically lead by the limited size of the data sample(s) in use.
The correct evaluation of the statistical uncertainty becomes more complex when splitting the dataset in phase-space regions and dealing with low numbers of candidates.

In particular, the variance on the denominator of $\varepsilon_\mathrm{Trig.}$ must be computed by decomposing its constituent parts into independent terms:
\begin{equation}
    \label{eqn:total-yield}
    \varepsilon_\mathrm{Trig.} = \frac{N_\mathrm{Trig.}}{\sum\limits_i{ \frac{\left(\alpha^i + \gamma^i\right)\left(\beta^i + \gamma^i\right)}{\gamma^i}}},
\end{equation}
for terms $\alpha^i=N_\mathrm{TIS}^i - N_\mathrm{TISTOS}^i$, $\beta^i = N_\mathrm{TOS}^i-N_\mathrm{TISTOS}^i$ and $\gamma^i = N_\mathrm{TISTOS}^i$, which contain exclusively \tis, \tos and \tistos candidates, respectively.
As demonstrated in \cite{LHCb-PUB-2014-039}, the variance of the denominator, $\sigma_{N_\mathrm{Tot.}}^2$ can be expressed as 
\begin{align*}
        \sigma_{N_\mathrm{Tot.}}^2 = &
        \sum\limits_i{\sigma_{N_\mathrm{Tot.}^i}^2}  \\ 
        \label{eqn:total-uncertainty}
                                  = & \sum\limits_i
            \left(\frac{\beta^i + \gamma^i}{\gamma^i}\right)^2 \sigma_{\alpha^i}^2 + \\ &
            \left(\frac{\alpha^i + \gamma^i}{\gamma^i}\right)^2 \sigma_{\beta^i}^2 +  \left(1 - \frac{\alpha^i \beta^i}{\left(\gamma^i\right)^2}\right)^2  \sigma_{\gamma^i}^2.
    \numberthis
\end{align*}
Note that the variances $\sigma_{\alpha^i/\beta^i}^2$ are defined such that ${\sigma_{N_\mathrm{TIS/TOS}^i}^2 = \sigma_{\alpha^i/\beta^i}^2 + \sigma_{N_\mathrm{TISTOS}^i}^2}$.

The statistical uncertainty on $\varepsilon_\mathrm{trig.}$ is determined by means of a generalised Wilson interval, as defined in \cite{Dembinski:2022}.
Defining an efficiency, $\hat{\varepsilon}$,  in terms of ``pass'' and ``fail'' counts, $m_1$ and $m_2$, the variance on each count can be expressed as
\begin{equation}
    \sigma_{m_i}^2 = m_i + \sigma_{i,b}^2,
\end{equation}
wherein $m_i$ is a Poisson contribution and $\sigma_{i,b}^2$ is a non-Poisson contribution.
The generalised Wilson interval given in \cite{Dembinski:2022} differs from the conventional Wilson interval~\cite{Wilson:1927xyh} by incorporating the contributions $\sigma_{1,b}^2$ and $\sigma_{2,b}^2$.

Assuming that $\hat{n}_i$ describes $n_i$ well, the contributions $\sigma_{i,b}^2$ can be computed as
\begin{subequations}
    \begin{equation}
        \sigma_{1,b}^2 = \sigma_{N_\mathrm{Trig.}}^2 - N_\mathrm{Trig.},
    \end{equation}
    \begin{equation}
        \sigma_{2,b}^2 = \sigma_{N_\mathrm{Tot.}}^2 - N_\mathrm{Tot.},
    \end{equation}
\end{subequations}
where $N_\mathrm{Tot.}$ and $\sigma_{N_\mathrm{Tot.}}^2$ are taken from Eqs.~\ref{eqn:total-yield}~and~\ref{eqn:total-uncertainty}, respectively.

\subsection{Systematic uncertainties}
\label{sec:uncertainties/systematic}

Systematic uncertainties may be assigned to trigger efficiencies estimated with the \tistos method to account for the effect of choices of:
\begin{itemize}
    \item the channel of interest, typically referred to as a calibration mode, in which the efficiencies are computed,
    \item the phase-space variables and regions scheme,
    \item the trigger decisions used to select the tag sample, 
    \item the background mitigation method.
\end{itemize}

The calibration mode should be chosen to be as representative as possible of the signal channel studied, \ie, topologically and kinematically similar, such as choosing \BToKll (where $\ell \in \left\{e, \mu\right\}$) as a calibration mode for \BToKllrare.
This similarity can be improved by employing reweighting techniques, such as in \cite{Rogozhnikov:2016bdp}.
A systematic uncertainty can be assigned by estimating trigger efficiencies for multiple calibration modes and taking the difference in the resulting efficiencies.

The choice on how the dataset is partitioned in regions of the decay phase space should balance a minimal statistical uncertainty (by choosing sufficiently large regions) and a minimal bias (by avoiding large regions as per Sec.~\ref{sec:efficiencies}).
To evaluate a systematic uncertainty for the choice of binning, analysts must compute efficiencies under varied phase-space divisions, \eg, by employing a binning scheme containing half or double as many bins.
The choice of phase-space variables can be varied similarly to assign a corresponding systematic uncertainty.

The trigger decisions chosen to select the tag sample in the \tistos method may lead to differing correlations between the tag and probe samples.
Whilst these correlations are mitigated when computing efficiencies in progressively smaller regions of the phase space, as discussed in Sec.~\ref{sec:efficiencies/tistos}, this mitigation may require smaller regions for certain tag decision choices.
According systematic uncertainties can be assigned by varying the choice of tag trigger decisions while keeping the same phase-space division scheme.

The choice of background mitigation method may lead to subtly different results on the evaluation of the trigger efficiencies.
No such differences were observed in Sec.~\ref{sec:triggercalib} for the \BToKmumu simulated sample; however, these samples are characterised by low-background and simple fit models.
When studying other decay channels, the assumptions made during these studies may no longer be valid and thus require further validation.
Analysts can compute systematic uncertainties on the trigger efficiency evaluation by varying the background mitigation method used and comparing the resulting efficiencies

\section{Conclusion}
\label{sec:conclusion}

The \tistos method of fully data-driven trigger efficiency estimation provides a robust framework through which to evaluate trigger efficiencies for the analysis of \bquark-decays.
A centralised implementation of this framework is provided in the \triggercalib software package, preventing the need for analysts to reimplement the same calculations in each physics analysis.
The package implements three methods of background mitigation and a novel approach to propagating statistical uncertainties, and enables analysts to apply corrections to the trigger response of simulated samples.
This functionality is demonstrated using Monte Carlo simulated ${\BToKmumu}$ decays in conditions equivalent to \lhcb data-taking in 2024, with the background mitigation approaches producing consistent results in these samples.
A discussion on the estimation of statistical and systematics uncertainties is given, with the latter highly dependent on the specifics of the analysis being performed, though where generally the correlations between trigger response and signal kinematics form the leading systematic uncertainty.

\begin{acknowledgements}
We would like to extend our sincere gratitude to the LHCb Real Time Analysis project for its support, for many useful discussions, and for reviewing an early draft of this manuscript.
We are grateful to the LHCb computing and simulation teams for producing the simulated LHCb samples used in the development of the method and package, and in their demonstration in this manuscript.
We would also like to thank our LHCb colleagues who have been involved in the development, implementation and validation of the methods and techniques described in this manuscript.

We acknowledge funding from the European Union Horizon 2020 research and innovation programme, call H2020-MSCA-ITN-2020, under Grant Agreement n. 956086.
This work has been sponsored by the German Federal Ministry of Education and Research (BMBF, grant no. 05H24PE2) within ErUM-FSP T04.
We also acknowledge the support of the German Academic Exchange Service received through the RISE Germany exchange scheme.
\end{acknowledgements}

\appendix

\section{Background model}
\label{sec:combinatorial}

To provide a background to the MC generated sample of \BToKmumu decays, toy candidates were generated.
This background component was constructed to mimic a typical combinatorial background, with an exponential shape in the \jpsi\Kp invariant mass and \Kp\mup\mun transverse momentum distributions.
This was achieved by drawing from exponential distributions with exponents of ${\ell_m = 5\times 10^{-3} \left(\!\mevcc\right)^{-1}}$ for $m\bigr(\jpsi\Kp\bigr)$ and ${\ell_{p_T} = 2\times 10^{-4} {\left(\!\mevc\right)}^{-1}}$ for $\pt\bigl(\Kp\mup\mun\bigr)$.

Taking the fraction of candidates in the simulated \BToKmumu sample with a \tis/\tos decision for each \hltone line, $i$, as $f^i_{\tis(\tos)}$, each candidate in the background sample was assigned a \tis/\tos decision based on a number drawn from a uniform random variable on $n = \left[0, 1\right]$.
The candidate was labelled \tis for a given line if $n < f^i_{\tis}$ and \tos if $n < f^i_{\tos} / 2$, with $n$ redrawn for each decision.

\section{\textit{sWeight} factorisation tests}
\label{sec:factorisation}

The likelihood ratio test divides a sample into two regions in the control variable and tests a null hypothesis (components have the same shape in both subsamples) and an alternate hypothesis wherein the shape of components of the distribution depend on the control variable.
This is tested by performing two likelihood fits: one ($H_0$) fitting a model simultaneously to both subsamples which shares shape parameters and the other ($H_1$) fitting independent PDFs to each subsample. From the resulting likelihoods, $L_{H_0}$ and $L_{H_1}$, respectively, a $Q$-statistic can be defined:
\begin{equation}
    Q = -2 \cdot (\ln\mathrm{sup}\{L_{H_0}\} - \ln\mathrm{sup}\{L_{H_1}\}).
    \label{eq:likelihood_test}
\end{equation}
A $p$-value is then obtained from a $\chi^2$ distribution with $N^{H_1} - N^{H_0}$ degrees of freedom, evaluated at the $Q$-statistic value.

Kendall's $\tau$ test takes an alternative approach, using pure signal and background samples (which can typically be obtained from signal simulation and sideband data subsamples) for the components present in the discriminating distribution.
For each subsample, the Kendall rank correlation coefficient, $\tau$, is computed between a control variable and the discriminating variable.
This $\tau$ is used to perform a hypothesis test wherein the null hypothesis is that the variables are independent, \textit{i.e}, $\tau = 1$.
As for the likelihood ratio test, a $p$-value is obtained from the hypothesis test which can be used to acce pt/reject the null hypothesis to a given confidence.
The test is only passed if the null hypothesis holds for both the signal and background samples. 

The likelihood test described in Sec.~\ref{sec:triggercalib} was performed on the \BToKmumu simulated sample with generated background, performing fits to the $\jpsi\Kp$ invariant mass and dividing the sample equally in $\pt\left(\Kp\mup\mun\right)$.
The mean and width of the signal and exponent of the combinatorial were shared between subsamples in the $H_0$ case and floated separately in the $H_1$ case.
Their values, the yields in each subsample and the corresponding minimised negative-log likelihood values are listed in Table~\ref{tab:factorisation}.
These yield a $Q$-statistic of $269.3$, corresponding to a $p$-value of $4.3\times10^{-58}$ for the 3 degrees of freedom differing between $H_0$ and $H_1$.
The null hypothesis $H_0$, that the $\jpsi\Kp$ invariant mass and $\pt\left(\Kp\mup\mun\right)$ are independent, is therefore rejected.
This conclusion is particularly evident when comparing the signal widths between the low- and high-\pt fits for $H_0$, where these differ by $\left(0.497 \pm 0.022\right)\mev$.

The Kendall $\tau$ test was also performed, taking MC simulated \BToKmumu events as the signal sample and candidates from the background generated according to App.~\ref{sec:combinatorial} as the background sample.
The results of this test are listed in Table~\ref{tab:kendall_tau}.
For a confidence of 99.7\%, the invariant mass and transverse momentum can only be considered independent in the background sample.

\begin{table*}[htb]
    \centering
    \caption{
        Results of the likelihood-ratio factorisation test.
    }
    \label{tab:factorisation}
    \begin{tabular}{ccccc}
        \toprule
        \multirow{2}{*}{Quantity} & \multicolumn{2}{c}{Simultaneous fit, $H_0$} & \multicolumn{2}{c}{Separate fits, $H_1$} \\
            & Low \pt & High \pt & Low \pt & High \pt  \\
        \midrule
        Signal mean, $\mu$ / \mev &
            \multicolumn{2}{c}{$5279.460 \pm 0.011$} &
            $5279.480 \pm 0.016$ &
            $5279.450 \pm 0.016$ \\
        Signal width, $\sigma$ / \mev &
            \multicolumn{2}{c}{$7.365 \pm 0.011$} &
            $7.586 \pm 0.015$ &
            $7.089 \pm 0.016$ \\
        Exponent, $\ell$ / $10^{-3} \mev^{-1}$ &
            \multicolumn{2}{c}{$\left(-4.932 \pm 0.016\right)\times 10^{-3}$} &
            $\left(-4.901 \pm 0.023\right)\times 10^{-3}$ &
            $\left(-4.966 \pm 0.021\right)\times 10^{-3}$ \\
        \midrule 
        Signal yield &
            $331130 \pm 660$ &
            $384320 \pm 700$ &
            $386920 \pm 670$ &
            $328050 \pm 700$ \\
        Combinatorial yield &
            $349710 \pm 670$ &
            $296520 \pm 630$ &
            $352790 \pm 690$ &
            $293922 \pm 640$ \\
        \midrule
        NLL, $-2\ln\sup\left\{L_{H_i}\right\}$ & \multicolumn{2}{c}{-9173670.41} &  \multicolumn{2}{c}{-9188431.25} \\
        $Q$-statistic & \multicolumn{4}{c}{269.3} \\
        $p$-value & \multicolumn{4}{c}{$4.3\times10^{-58}$} \\
        \bottomrule
    \end{tabular}
\end{table*}
    
\begin{table*}[htb]
    \centering
    \caption{
        Results of the Kendall $\tau$ test.
    }
    \label{tab:kendall_tau}
    \begin{tabular}{ccc}
        \toprule
        Quantity & Signal (MC simulation) & Background (per App.~\ref{sec:combinatorial}) \\
        \midrule
        $\tau$ coefficient &  $\left(1.151 \pm 0.077\right)\times 10^{-2}$ & $\left(6.4 \pm 7.2\right)\times 10^{-3}$ \\
        $p$-value & $ 1.1 \times 10^{-50} $ & $0.37$ \\
        \bottomrule
    \end{tabular}
\end{table*}

\bibliographystyle{spphys}       
\bibliography{main, standard, LHCb-CONF, LHCb-DP, LHCb-PAPER, LHCb-TDR}   


\end{document}